\documentclass{article}

\usepackage[nonatbib,final]{neurips_2024}

\usepackage[utf8]{inputenc} % allow utf-8 input
\usepackage[T1]{fontenc}    % use 8-bit T1 fonts
\usepackage{url}            % simple URL typesetting
\usepackage{amsfonts}       % blackboard math symbols
\usepackage{nicefrac}       % compact symbols for 1/2, etc.
\usepackage{microtype}      % microtypography
\usepackage{xcolor}         % colors

\usepackage[style=ieee,citestyle=numeric-comp,backend=biber,natbib=true,sorting=none,maxbibnames=10,mincitenames=1, minbibnames=1]{biblatex}

\usepackage[hidelinks]{hyperref}       % hyperlinks
\usepackage{xurl} % fix biblatex's URL wrapping
\hypersetup{breaklinks=true}

\AtEveryBibitem{\clearfield{urlyear}\clearfield{note}} %remove undesired fields
\DefineBibliographyStrings{english}{andothers = {\upshape et al\adddot}} %un-italicize "et al."

\usepackage[sort,capitalize]{cleveref}

\title{On the Ethical Considerations of Generative Agents}

\author{%
  N'yoma Diamond \\
  Department of Computer Science and Technology \\
  University of Cambridge \\
  Cambridge, England \\
  \texttt{bad35@cantab.ac.uk} \\
  \And
  Soumya Banerjee \\
  Department of Computer Science and Technology \\
  University of Cambridge \\
  Cambridge, England \\
  \texttt{sb2333@cam.ac.uk}
}

\begin{document}

\maketitle

\begin{abstract}
    The Generative Agents framework recently developed by \citeauthor{park_generative_2023} has enabled numerous new technical solutions and problem-solving approaches. Academic and industrial interest in generative agents has been explosive as a result of the effectiveness of generative agents toward emulating human behaviour. However, it is necessary to consider the ethical challenges and concerns posed by this technique and its usage. In this position paper, we discuss the extant literature that evaluate the ethical considerations regarding generative agents and similar generative tools, and identify additional concerns of significant importance. We also suggest guidelines and necessary future research on how to mitigate some of the ethical issues and systemic risks associated with generative agents. % perspective piece
    % substantial theorised applications, and the simultaneous meteoric interest in generative artificial intelligence (GAI).
    % New uses for generative agents are being developed every day as researchers and developers apply the framework to more problems. 
\end{abstract}

%\section{A Brief Overview of Generative Agents}
\section{Introduction}

% \TODO{discuss issues relevant to broader GAI may be specifically exacerbated by generative agents (risks are not exclusive)}
% \TODO{discuss identified concerns are not exhaustive; collated and simplified from identified literature, plus additional concerns proposed by authors needing greater discussion.}

The seminal work by \citet{park_generative_2023} introduced the Generative Agents framework for simulating human behaviour using generative language models. Generative agents have the ability to operate independently and creatively, making decisions to reach a goal with minimal user input. To do this, each agent maintains a distinct and dedicated set of memories and behavioural parameters that are provided to a generative language model as needed. By recording memories, prompting reflection on them, planning future actions, and reacting to short-term observations, generative agents can produce realistic emulations of human behaviour and decision-making~\cite{park_generative_2023}. As a result, generative agents have massive potential for widespread adoption and may revolutionise many modern problems. However, this introduces a number of critical ethical challenges regarding their research, development, and application. In this position paper, we examine the existing literature on the ethical aspects of generative agents, highlight a handful of critical ethical concerns, and discuss recommendations for their mitigation. We note that the concerns posed in this work are not exhaustive or entirely exclusive to generative agents, but are believed to be especially important to discuss in this specific context.

\section{Existing Literature}

\citet{bail_can_2024} provides a detailed qualitative analysis and discussion of the potential benefits of generative artificial intelligence (GAI) and generative agents towards social science. With respect to ethical considerations, \citet{bail_can_2024} primarily focuses on GAI over generative agents specifically. \citet{bail_can_2024} identifies that GAI tools are highly limited by their reliance on human-generated knowledge and are thereby vulnerable to exhibit human biases. They also question whether GAI can produce high-quality realistic and accurate research which is truly actionable, particularly informed by many existing GAI technologies' propensity to hallucinate and produce misinformation. Finally, \citet{bail_can_2024} raises concerns over the reproducibility of research using GAI due to its inherent stochasticity and the massive quantity of non-peer-reviewed GAI-based literature being produced, potentially harmfully contributing to ongoing replication crises~\cite{ioannidis_why_2005,shrout_psychology_2018,wiggins_replication_2019,tackett_psychologys_2019}. However, \citet{bail_can_2024} also identifies the critical opportunities for open social science research posed by GAI, such as through the development and improvement of open-source GAI tools and infrastructure, citing precedent from the (former) open sharing of research data generated by many social media companies. 

\citet{lazar_frontier_2024} considers a narrative rhetoric approach towards analysing the potential societal impacts of generative agents. In their work, \citet{lazar_frontier_2024} identifies prior challenges produced by GAI, their relevance towards recent developments in GAI, and existing approaches' failure to overcome some of these hurdles. They also discuss specific applications of generative agents and the ethical questions and concerns they face from the perspective of societal impact: Generative agent-based companions can provide social engagement, assist in day-to-day tasks, improve recommender and information retrieval systems, and serve as ``universal intermediaries'' by consolidating and automating useful tasks and jobs. However, this may cause users to develop unhealthy parasocial relationships and dependencies, displace human labour, commodify human attention, and worsen user privacy and security~\cite{lazar_frontier_2024}. Notably, while \citet{lazar_frontier_2024} adequately identifies these numerous challenges, they do some from a primarily rhetorical standpoint. As a result, they neglect to suggest means to address many of the stated concerns. 

\citet{chan2023harms} provide an overview and analysis of future harms of agentic algorithmic systems more broadly. By focusing on agentic algorithmic systems as a broad category (including generative agents, among other technologies), \citet{chan2023harms} justify the need to anticipate future harms of such systems, identify specific examples, and propose approaches and areas of future work to explore and implement to prevent the identified harms before they occur. Notably, while they do make note of GAI-based tools for specific examples, they are not a specific focus of the work. As a result, many risks and challenges specific to agentic GAI-based tools, like generative agents, are ignored.

\citet{anwar2024foundational, gabriel2024ethicsadvancedaiassistants} provide comprehensive and in-depth discussion of the technical challenges and ethical risks of GAI-based systems, respectively. Both works specifically consider generative agents or similar systems, making them the most directly related extant literature to this work. \citet{anwar2024foundational} primarily focus on the technical challenges of GAI-based systems and how flaws in their design, training, tuning, and implementation may lead to critical problems. Specifically, in their discussion about agentic LLMs and multi-agent systems, \citet{anwar2024foundational} identify risks resulting from the combination of generative natural language models and agentic behaviour, such as difficulty ensuring desired behaviour, providing robust oversight, and ensuring safe behaviour when afforded access to external abilities and services. The authors also identify risks associated with the emergent behaviours of multi-agent systems and the lack of extant research into understanding them directly. Based on these and other identified challenges, \citet{anwar2024foundational} discuss a series of sociotechnical challenges, many of which have ethical implications in line with those discussed in this work. \citet{gabriel2024ethicsadvancedaiassistants} specifically focus on what they refer to as ``AI assistants'', which are GAI-based agents capable of independent and autonomous planning and action on behalf of a user via text-based instruction. This case is a specific example within the broader category of generative agents, however it serves as an effective case-study towards identifying the risks and ethical concerns of generative agents. Using this lens, \citet{gabriel2024ethicsadvancedaiassistants} identify a series of significant personal, socioeconomic, and environmental concerns posed by AI assistants, many of which we discuss in this work.

\section{Ethical Concerns of Generative Agents}

\subsection*{Anthropomorphisation and Misunderstanding of Experimental Results}

Anthropomorphisation can serve as an effective means to discuss and rationalize the behaviour of AI tools, such as generative agents. However, generative agents and other GAI tools as they exist today do not actually possess any level of consciousness and are incapable of producing true emotion or intention. Thus, a problematic disconnect can develop between how AI tools are understood, discussed, and utilised as compared to the reality of their capabilities. This poses a significant ethical concern for the usage and understanding of generative agents with respect to human behaviour. Present literature provides substantial discussion of anthropomorphisation of generative agents and similar tools towards users in social applications (such as chatbots)~\cite{lazar_frontier_2024,gabriel2024ethicsadvancedaiassistants,abercrombie_mirages_2023}. However, no extant research has been identified directly analysing the impacts and risks associated with how results generated using generative agents may be erroneously interpreted due to excessive anthropomorphisation. That is, prescribing undue anthropomorphic characteristics to generative agents risks understating its purpose as an emulator of human behaviour, and not a direct model of how humans actually behave.

A critical benefit of generative agents is the potential to serve as an effective tool to perform human-like actions where humans cannot be used, such as to emulate and understand human behaviour or complex human-interaction systems. However, excessive anthropomorphisation risks critically misunderstanding the results of such experiments and creating misinformation. That is, any result produced using generative agents is only descriptive of the behaviour of the implemented framework and its underlying generative model. The behaviours of generative agents and language models are not causally linked to human behaviour patterns. To ensure the quality and accuracy of research conducted using generative agents, conscious effort must be made to avoid misattributing the characteristics of generative agents towards the behaviour of real humans. In addition, more research must be conducted to understand the risks associated with such distorted interpretations.

\subsection*{Creation of Parasocial Relationships}

Anthropomorphisation also risks the creation of parasocial relationships between AI and its users~\cite{lazar_frontier_2024, gabriel2024ethicsadvancedaiassistants}. This is particularly true for generative agents, which may be individualised and/or physically embodied in future applications, such as explored by \citet{gabriel2024ethicsadvancedaiassistants}. The structure of generative agents lend themselves towards the creation of individualised agents which non-technical users may develop relationships. Individualised or embodied generative agents may be an effective tool toward solving day-to-day problems---much like existing virtual assistants such as Siri and Google Assistant. However, the persona, memory, and reflection characteristics of generative agents allow them to be perceived as more similar to humans than existing virtual assistants despite not being fundamentally any more ``intelligent''. As a result, the users of such agents risk harmfully anthropomorphising GAI and developing harmful relationships or attachments to these tools.

While \citet{gabriel2024ethicsadvancedaiassistants} discuss ways to mitigate risks induced by parasocial relationships, they mainly focus on post hoc utilitarian and material approaches to optimise reliable agent performance in cases of human-AI relationships, rather than avoiding them altogether. \citet{park_generative_2023,abercrombie_mirages_2023} suggest that generative agents and other GAI tools should be designed to explicitly state their nature as generative models and directly avoid anthropomorphic language and behaviour. However, little consideration has yet to be given to how this may be reliably implemented and the effectiveness of such approaches. In particular, many of these methods burden the user with identifying and mitigating parasocial relationships. Such approaches may prove unhelpful or even counterproductive if users ignore, misinterpret, or intentionally circumvent the provided warnings or guardrails~\cite{anwar2024foundational}. Further, anthropomorphic characteristics are not objectively identifiable and the boundaries distinguishing anthropomorphic behaviour are becoming exceedingly unclear. As a result, approaches to reliably avoid anthropomorphic behaviours will be extremely difficult to develop.

As a simple example, the implementation provided by \citet{park_generative_2023} for their generative agent simulation framework defers to the underlying generative model to identify unsafe behaviour. Upon query, the generative model is prompted to score how much a user's prompt anthropomorphizes the agent. If the score is greater than a critical threshold, then the user is told that attributing human agency to generative agents is inappropriate and their query is rejected. This aims to ensure that the framework is being used safely. However, this method is very simplistic and relies on the untested assumption that the underlying model is capable of accurately identifying harmful anthropomorphisation. In reality, the ability to identify such characteristics will vary greatly between models and is not provably accurate. In addition, this system provides no guardrails to prevent the user from ignoring or intentionally circumventing these checks.

\subsection*{Excessive Trust and Insufficient Scepticism}

Similar to concerns regarding the misattribution of human characteristics, over-reliance and overconfidence in generative agents may result in the unintentional spread of misinformation. Specifically, users may prescribe undue trustworthiness to generative agents~\cite{choudhuryInvestigatingImpactUser2023,zhanDeceptiveAIEcosystems2023,gabriel2024ethicsadvancedaiassistants,anwar2024foundational}. Intuitively, generative agents should be more capable than existing technologies (such as AI chatbots or virtual assistants) due to their memory and reflection modules, as these modules allow generative agents to better recall information and develop critical insights~\cite{park_generative_2023}. However, generative agents are still just an extension of existing language models and thus are prone to the same mistakes and errors. Such errors include hallucinations, poisoning, failure to recall available information, or recapitulating and reinforcing endemic biases. Further, the stochasticity of generative language models can induce unreliable and/or inaccurate behaviour from generative agents~\cite{bender2021stochastic,bail_can_2024}.

A lack of critical analysis and scepticism of responses produced by generative agents runs the risk of unintentionally spreading misinformation and reinforcing harmful biases~\cite{choudhuryInvestigatingImpactUser2023,zhanDeceptiveAIEcosystems2023, gabriel2024ethicsadvancedaiassistants,anwar2024foundational}. \citet{gabriel2024ethicsadvancedaiassistants} specifically discuss the angle of (misplaced) trust and misinformation associated with AI assistants. In particular, they identify varying types of trust that a user may have in GAI-based systems like generative agents, such as overestimating the competence of an agent or the quality of its alignment. They also point out that undue trust makes people highly vulnerable to misinformation, manipulation, and ideological entrenchment---dynamics highly similar to those observed in humans, and potentially even more concerning if generative agents face widespread adoption. To address these risks, \citet{gabriel2024ethicsadvancedaiassistants} collate a handful of technical design and policy-based proposals, such as identifying and indicating uncertainty, minimising (un)necessary complexity, and improving technical transparency and public understanding of GAI systems. However, many of these approaches require severely limiting desirable agent functionality or leveraging external response analysis methods which are vulnerable to circumvention. Thus, they may only serve as stop-gap solutions. As a result, there is still a significant need for further research and development of techniques that can inherently mitigate these concerns.

Ideally, internal technical mechanisms should be developed to provably ensure that provided information is true and impartial, or otherwise marked as uncertain or potentially biased. To this end, some researchers have suggested that retrieval-augmented generation (RAG) techniques may provide a technical solution to mitigate generative agents spreading or unintentionally creating misinformation by leveraging access to external knowledge resources during inference~\cite{chenBenchmarkingLargeLanguage2024,ramInContextRetrievalAugmentedLanguage2023,fanSurveyRAGMeeting2024}. However, we were unable to identify any literature applying these techniques to memory-/reflection-enabled generative agents as proposed by \citet{park_generative_2023}. Further, RAG simply provides more reliable access to information and does not inherently prevent them from producing misinformation. It is possible for RAG-based agents to produce erroneous inferences that are only partially informed by retrieved information. That is, RAG-enabled generative agents may draw incorrect conclusions when provided with incomplete or highly complex knowledge. In addition, information retrieval can itself unintentionally provide the model with misinformation under certain circumstances, such as via design error or scope misalignment. This can cause even greater harm due to the elevated credibility that would likely be attributed to RAG-enabled generative agents~\cite{chenBenchmarkingLargeLanguage2024,fanSurveyRAGMeeting2024}.

\subsection*{Usage by Malicious Actors}

The prevalence of automated bots has become a key driver in the spread of misinformation online, significantly harming access to trustworthy and accurate information~\cite{himelein-wachowiak_bots_2021,hajli_social_2022,shao_spread_2018,mirsky_creation_2021,lim_fact_2023,anwar2024foundational,gabriel2024ethicsadvancedaiassistants}. Simultaneously, GAI tools such as ChatGPT are beginning to be applied toward the creation of automated scams and phishing attacks~\cite{na_evolving_2023,roy_chatbots_2024,gabriel2024ethicsadvancedaiassistants,anwar2024foundational}. As automated means for conducting malicious acts become more prevalent, generative agents may be particularly susceptible to misuse. Malicious actors can leverage generative agents as automated tools to spread disinformation, execute scams, or conduct cyberattacks. The distinct memory and reflection characteristics of generative agents make them capable of performing malicious actions (such as spreading misinformation or conducting scams) with greater realism and effectiveness than existing technologies. Thus, malicious generative agents may be better at deceiving humans and avoiding automated detection than existing approaches. \citet{gabriel2024ethicsadvancedaiassistants} briefly discuss the types of security vulnerabilities GAI systems introduce to users, and how malefactors' abilities may be enhanced by GAI systems leveraging memory, planning, and reflections capabilities like those proposed by \citet{park_generative_2023}. However, they do not consider specific concerns posed by automated generative agents.

Further, malicious generative agents may perform social engineering attacks, impersonate friends, relatives, or officials, and conduct other cyberattacks that were previously only possible by humans~\cite{alotaibi_cyberattacks_2024,ferrara_genai_2024}. To this end, future work should consider developing techniques to detect the usage of automated generative agents~\cite{gabriel2024ethicsadvancedaiassistants}. However, such work is easier said than done, as automatically and accurately distinguishing and mitigating malicious behaviour is exceptionally difficult. In practice, it may not be effective to specifically target GAI in lieu of broader misinformation, scam, or attack detection. Thus, additional practical solutions also need to come from legislating acceptable development and usage of generative agents to suppress misuse~\cite{gabriel2024ethicsadvancedaiassistants,anwar2024foundational}.

\subsection*{Vulnerability to Hijacking}

In addition to direct usage by malicious actors, developers of generative agents must be wary of their vulnerability to hijacking or jailbreaking. GAI tools are prime targets for attacks that aim to derail system behaviour. This is due to their usage in a wide variety of applications, the transferability of attacks between implementations, and the difficulty of developing effective behavioural guardrails~\cite{chowdhuryBreakingDefensesComparative2024,xuComprehensiveStudyJailbreak2024}. As with previously discussed concerns, generative agents' memory and reflection capabilities make them highly desirable and effective for an increasing range of tasks, raising direct concerns about hijacking as they see greater adoption. Hijacked generative agents may be difficult to recover, have access to more sensitive information and actions than other GAI tools, and potentially automatically hijack other agents or systems like a worm virus. This is especially concerning as substantial research has already been conducted towards developing techniques to compromise generative language models~\cite{chowdhuryBreakingDefensesComparative2024,greshakeNotWhatYou2023,kilovatyHackingGenerativeAI2024,leviVocabularyAttackHijack2024,xuComprehensiveStudyJailbreak2024}. Notably, the desire to hijack GAI systems is not exclusive to malicious actors, as normal users may also directly benefit from hijacking (or ``jailbreaking'') automated AI-based tools~\cite{notopoulosCarDealershipAdded2023,debterRetailersAreTesting2023,faithfullFutureHagglingBargain2024}. Thus, hijacking and jailbreaking serve both as a threat vector for external actors to harm users of generative agents and for users to circumvent their safety features~\cite{gabriel2024ethicsadvancedaiassistants,anwar2024foundational}. 

In response to these concerns, many authors suggest a need to improve approaches to detect, understand, and mitigate agent hijacking~\cite{gabriel2024ethicsadvancedaiassistants,anwar2024foundational}. However, we believe a simpler approach should be considered with greater interest; critical assessment of the usage of generative agents altogether. Given the difficulty of detecting and preventing model hijacking and our as-yet lacking understanding of these problems~\cite{anwar2024foundational}, developers must be wary of where, when, how, and if they should implement generative agents at all. In particular, generative agents should not be utilised in contexts where their hijacking may enable significant threat vectors, such as those suggested by \citet{greshakeNotWhatYou2023,gabriel2024ethicsadvancedaiassistants,anwar2024foundational}. In cases where it may not be possible to eliminate the usage of vulnerable generative agents, safeguards must be developed to prevent agents from being actionable when hijacked. For example, sensitive information or services may require secondary human authentication to be accessed, or generative agents may be sandboxed to prevent a hijacked agent from spreading undesirable behaviour to other agents or systems.

\subsection*{Displacement of Human Labour}

Discussion of the ethical concerns of any technology would be incomplete without analysis of its effect on human labour. This is especially true for generative agents, as automated generative agents may be highly attractive to organisations that wish to automate tasks that are generally performed by humans. Generative language models have already begun to replace humans in both low- and high-skill occupations, such as customer service agents~\cite{verma_chatgpt_2023}, translators~\cite{yilmaz_ai-driven_2023}, and varying forms of knowledge experts~\cite{yilmaz_ai-driven_2023,kalla_study_2023}. As a result, generative agents are expected to have significant impact on the quantity and quality of human labour, among other socioeconomic factors~\cite{gabriel2024ethicsadvancedaiassistants,anwar2024foundational}

Currently, AI tools are still not sophisticated enough to fully replace humans in many positions~\cite{farhi_news_2023}. However, the design benefits of generative agents potentially introduce the necessary capabilities to be effectively leveraged in many of these applications. As a result, overzealous usage of generative agents may cause the rapid displacement of human workers across many occupations in a particularly disruptive and unprecedented manner~\cite{anwar2024foundational}. As such, organisations and researchers should prioritize techniques that use generative agents as collaborative tools to supplement human workers instead of attempting to replace human labour. This approach is validated by the extant literature, which asserts that human-AI collaboration is highly effective at improving productivity while simultaneously not harming the current human workforce~\cite{brynjolfsson_generative_2023,dellermann_future_2021,fui-hoon_nah_generative_2023,lai_human-ai_2021,sowa_cobots_2021}.

\subsection*{Exploitation of Developing Nations and Modern Slavery}

As demand grows for generative agents and other GAI-based tools, so will the necessity to manufacture physical hardware and electronic devices capable of leveraging them (such as GPUs and smartphones). Much of the natural and human resources used in the manufacture of these devices comes from developing nations, where there are significant risks of exploitation and contributing to modern slavery~\cite{von_der_goltz_mines_2019,maier_socially_2014,dorninger_global_2021,han_modern_2022}. Notably, we were unable to find any literature providing meaningful discussion of these concerns as they relate to GAI, despite substantial documentation of the impacts of the physical resources they require. In particular, the mining of silica---a primary component of computer chips---and lithium---a primary component of batteries---and other materials used to manufacture these electronic devices can pose significant environmental and personal health risks for miners and their communities~\cite{mishra_impact_2015,ribeiro_need_2021,maier_socially_2014,vidal_how_2015,von_der_goltz_mines_2019}.

Individuals and organisations wishing to develop and use generative agents or similar tools must ensure that they are not contributing to these risks. This can be done by evaluating the hardware requirements associated with using generative agents and minimising them wherever possible. Future work should aim to improve the efficiency of generative agents and the required hardware to reduce the need for materials and components whose manufacture and usage may contribute to exploitation and modern slavery. Further, independent audits should be conducted on manufacturers and other sections of the supply-chain to ensure adherence to these principles~\cite{han_modern_2022}. Finally, requirements for such hardware should be avoided wherever possible, such as through the using simpler and more sustainable techniques or eliminating superfluous usage of generative agents.

\subsection*{Environmental Impact}

Alongside increased demand for AI-based tools, the environmental impact and effective carbon footprint of GAI systems have also increased substantially~\cite{dhar_carbon_2020,wu_sustainable_2022,samsi_words_2023,jiang_preventing_2024,gabriel2024ethicsadvancedaiassistants}. \citet{gabriel2024ethicsadvancedaiassistants} identify a range of ways in which the creation and usage of AI assistants, a notable potential application of generative agents, can impact the environment. Specifically, high energy usage during training and inference, emissions embodied by the production of required hardware, and supporting environmentally irresponsible industries and applications all pose substantial environmental risk. Further, the nature and magnitude of embodied emissions associated with model training and the manufacture of required hardware are only just beginning to be investigated and understood~\cite{faiz_llmcarbon_2024,jiang_preventing_2024,ligozat_unraveling_2022,mulligan_ai_2022,gabriel2024ethicsadvancedaiassistants}. 

To mitigate these concerns, \citet{gabriel2024ethicsadvancedaiassistants} suggest a number of emissions-reduction approaches, such as minimising model size, improving hardware efficiency, sourcing carbon-free energy, and implementing public policy to encourage environmentally sustainable development and deployment. However, these approaches heavily rely on the accessibility and feasibility of low-emissions resources and techniques. As such, similar to the discussed concerns about hijacking, we believe a simpler and more feasible approach would be to minimise or eliminate the usage of generative agents wherever possible. Specifically, the implementation of generative agents must be critically analysed with respect to their necessity and weighed against their environmental impact to avoid superfluous or redundant usage~\cite{dhar_carbon_2020,ligozat_unraveling_2022,jiang_preventing_2024,mulligan_ai_2022}.

\section{Conclusion}

In summary, the development and application of generative agents present many ethical challenges and concerns. Ethical problems arise from all sides of GAI usage, including developers, malicious actors, and normal users. Frivolous implementation, unsafe or inefficient design, unreliable and untrustworthy behaviour, and user error all pose significant threats to the ethical usage of generative agents. Given these challenges, continued overzealous acceleration of generative agent development needs to be considered critically and addressed. Currently, the ethical evaluations and impacts of newly developed GAI and generative agent techniques are often left as an afterthought, if they are discussed at all. Thus, future research and applications of generative agents should directly account for the ethical concerns posed in this work and those identified in other works, such as \citet{gabriel2024ethicsadvancedaiassistants,anwar2024foundational,lazar_frontier_2024,bail_can_2024,chan2023harms}, among others. To this end, researchers, developers, and legislators should apply the proposed mitigation approaches wherever possible, and create new mitigation techniques where solutions have yet to be developed.

% \bibliography{references} %natbib method
\printbibliography %biblatex method

\end{document}